\documentclass[12pt]{article}
\usepackage{ctex}
\usepackage{amssymb, amsmath, color, mathdots}
\usepackage[colorlinks, linkcolor=blue]{hyperref}
\usepackage{geometry}
\usepackage{tabularx}
\usepackage{longtable}
\usepackage{caption}
\usepackage{threeparttable}
\usepackage{rotating}
\usepackage{bicaption}
\usepackage{bm}
\usepackage{authblk}
\usepackage{multirow}
\usepackage{makecell}
\usepackage[square, comma, sort&compress, numbers]{natbib}

\begin{document}
	\title{\bf The error-correcting pair for several classes of NMDS linear codes}
			
			

	


	\author{\small Dong He, Zhaohui Zhang}	
\author{\small Qunying Liao
	\thanks{Corresponding author.
		
		{~~E-mail. qunyingliao@sicnu.edu.cn (Q. Liao), 759803613@qq.com(D. He).}
		
		{~~Supported by National Natural Science Foundation of China (12471494) and Natural Science Foundation of Sichuan Province (2024NSFSC2051).}	}
}

	\affil[] {\small(College of Mathematical Science, Sichuan Normal University, Chengdu, 610066, China)}
	\date{}
	\maketitle
	\newtheorem{theorem}{Theorem}[section]
	\newtheorem{example}{Example}[section]
	\newtheorem{definition}{Definition}[section]
	\newtheorem{lemma}{Lemma}[section]
	\newtheorem{proposition}{Proposition}[section]
	\newtheorem{corollary}{Corollary}[section]
	\newtheorem{remark}{Remark}[section]
	\renewcommand\refname{References}	
	\renewcommand{\theequation}{\thesection.\arabic{equation}}
	\newcommand{\upcite}[1]{\textsuperscript{\textsuperscript{\cite{#1}}}}
	\catcode`@=11 \@addtoreset{equation}{section} \catcode`@=12
	{\bf Abstract.}  
	{\small  The error-correcting pair is a general algebraic decoding method for linear codes. The near maximal distance separable (NMDS) linear code is a subclass of linear codes and has applications in secret sharing scheme and communication systems due to the efficient performance, thus we focus on the error-correcting pair of NMDS linear codes. In 2023, He and Liao showed that for an NMDS linear code $\mathcal{C}$ with minimal distance $2\ell+1$ or $2\ell+2$, if $\mathcal{C}$ has an $\ell$-error-correcting pair $\left( \mathcal{A}, \mathcal{B} \right)$, then the parameters of $\mathcal{A}$ have 6 or 10 possibilities, respectively.
		In this manuscript, basing on Product Singleton Bound, we give several necessary conditions for that the NMDS linear code $\mathcal{C}$ with minimal distance $2\ell+1$ has an $\ell$-error-correcting pair $(\mathcal{A}, \mathcal{B})$, where the parameters of $\mathcal{A}$ is the 1st, 2nd, 4th or 5th case, then basing on twisted generalized Reed-Solomon codes, we give an example for that the parameters of $\mathcal{A}$ is the 1st case. Moreover, we also give several necessary conditions for that the NMDS linear code $\mathcal{C}$ with minimal distance $2\ell+2$ has an $\ell$-error-correcting pair $(\mathcal{A}, \mathcal{B})$, where the parameters of $\mathcal{A}$ is the 2nd, 4th, 7th or 8th case, then we give an example for that the parameters of $\mathcal{A}$ is the 1st or 2nd case, respectively.
	}
	

	\section{Introduction}
	Information security plays an essential role in security of data transmission and storage, which is critical for personal privacy, corporate secrets and national security. Coding theory provides powerful tools to ensure that the information can be accurately received and decoded.
	
	For a linear code $\mathcal{C}=[n, k, d]$, if 
	the Singleton Bound is reached, namely, $d=n-k+1$, then $\mathcal{C}$ is maximum distance separable (in short, MDS). 
	 MDS linear codes as a class of linear codes are very important in data storage and data transmission [\citealp{A11},\citealp{A6},\citealp{A24}]. A $q$-ary $[n,k,d]$ linear code $\mathcal{C}$ is almost maximum distance separable (in short, AMDS) if $d=n-k$. Especially, an AMDS code $\mathcal{C}$ is near maximum distance separable (in short, NMDS) if $\mathcal{C}^{\perp}$ is also AMDS. NMDS codes are closely related to design theory and finite geometry, and also have applications in secret sharing scheme [\citealp{A33}]
	 Recently, several types of NMDS codes were constructed  [\citealp{A27},\citealp{A28},\citealp{A26},\citealp{A29},\citealp{A30}]. 
	
	As a general algebraic method for decoding linear codes, the error-correcting pair was introduced by Kötter [\citealp{A2}] and Pellikaan [\citealp{A1}] in 1992, independently. In brief, 
	an $\ell$-error-correcting pair of $\mathcal{C}$ is a pair of linear codes,
	denoted by $(\mathcal{A}, \mathcal{B})$, which determines a decoding algorithm for $\mathcal{C}$ and can correct up to $\ell$ errors in polynomial times. The pair alreadly exists for several classes of linear codes, such as GRS codes, TGRS codes, cyclic codes, Goppa codes, alternant codes and algebraic geometry codes [\citealp{A15},\citealp{A4},\citealp{A38},\citealp{A6},\citealp{A5},\citealp{A3}].

	For an MDS linear code $\mathcal{C}$, in 2016, Márquez-Corbella and Pellikaan [\citealp{A6}] gave a necessary and sufficient condition for that $\mathcal{C}$ with minimal distance $2\ell+1$ has an error-correcting pair. In 2023, He and Liao \cite{A25} showed that if  $\mathcal{C}$ is with minimal distance $2\ell+2$ and has an $\ell$-error-correcting pair $\left( \mathcal{A}, \mathcal{B} \right)$, then the parameters of $\mathcal{A}$ have 3 cases, moreover, they gave a necessary condition  for the 1st case, and then Xiao and Liao  [\citealp{A32}] gave a necessary condition for the 2nd case. Recently, He and Liao [\citealp{A31}] proved that if an NMDS linear code $\mathcal{C}$ with minimal distance $2\ell+1$ or $2\ell+2$ has an $\ell$-error-correcting pair $\left( \mathcal{A}, \mathcal{B} \right)$, then there are 6 cases or 10 cases for the parameters of $\mathcal{A}$, respectively. In this manuscript, we give several necessary conditions for that the NMDS linear code $\mathcal{C}$ with minimal distance $2\ell+1$ has an $\ell$-error-correcting pair $(\mathcal{A}, \mathcal{B})$ when the parameters of $\mathcal{A}$ is the 1st, 2nd, 4th or 5th case, then we give an example for that the parameters of $\mathcal{A}$ is the 1st case basing on TGRS codes. Moreover, we give several necessary conditions for that the NMDS linear code $\mathcal{C}$ with minimal distance $2\ell+2$ has an $\ell$-error-correcting pair $(\mathcal{A}, \mathcal{B})$ when the parameters of $\mathcal{A}$ is the 2nd, 4th, 7th or 8th case, then we give an example for that the parameters of $\mathcal{A}$ is the 1st or 2nd case basing on TGRS codes.
	
	This manuscript is organized as follows. In Section 2, some necessary notions, definitions and lemmas are given. 
	In Section 3, the main results for that an NMDS linear code $\mathcal{C}$ with minimal distance $2 \ell+1$ has an $\ell$-error-correcting pair are presented.
	In Section 4, the main results for that an NMDS linear code $\mathcal{C}$ with minimal distance $2 \ell+2$ has an $\ell$-error-correcting pair are presented. In section 5, we give a brief proof for that an MDS code with minimal distance $2\ell+2$ and an $\ell$-error-correcting pair $(\mathcal{A}, \mathcal{B})$ is GRS, when the parameters of $\mathcal{A}$ is the 1st or 2nd case. In section 6, the conclusions of the whole manuscript are given.
	
	\section{Preliminaries\text}
	
	In this section, we give some necessary notations as follows.
	
	$\bullet$ $\mathbb{F}_q$ denotes the finite field of $q$ elements, where $q$ is a prime power.
	
	$\bullet$ $\mathbb{F}_{q}^m$ denotes the vector space with dimension $m$ over $\mathbb{F}_q$, where $m$ is a positive integer.
	
	$\bullet$ $\mathbb{F}_q[x]$ is the polynomial ring over $\mathbb{F}_q$.
	
	$\bullet$ Denote $[n]=\{1, 2, \ldots, n\}$, $\mathbf{1}=\left(1,\ldots,1\right)\in\mathbb{F}_q^n$ and $\mathbf{0}=\left(0,\ldots,0\right)\in\mathbb{F}_q^n$. 
	
	
	$\bullet$ For two vectors $\boldsymbol{x}=\left(x_1, \ldots, x_n\right)$ and $\boldsymbol{y}=\left(y_1, \ldots, y_n\right) \in \mathbb{F}_q^n$, the standard inner product and the Schur product of $\boldsymbol{x}$ and $\boldsymbol{y}$ are respectively defined as
	$$
	\boldsymbol{x} \cdot \boldsymbol{y}=\sum_{i=1}^n x_i y_i \quad \text { and } \quad \boldsymbol{x} * \boldsymbol{y}=\left(x_1 y_1, \ldots, x_n y_n\right) .
	$$
	
	$\bullet$ $k(\mathcal{C})$ and $d(\mathcal{C})$ denote the dimension and minimum (Hamming) distance of the linear code $\mathcal{C}$, respectively.
	
	$\bullet$ Let $U$, $V$ and $\mathcal{C}$ be $q$-ary subspaces of $\mathbb{F}_q^n$. Denote $\mathcal{C}^{\perp}=\{\boldsymbol{x} \mid \boldsymbol{x} \cdot \boldsymbol{c}= 0$, $\forall \boldsymbol{c} \in \mathcal{C}\}$ to be the dual code of $\mathcal{C}$. Furthermore, $U \perp V$ means that $\boldsymbol{u} \cdot \boldsymbol{v}=0$ for any $\boldsymbol{u} \in U$ and any $\boldsymbol{v} \in V$. According to the definition of the Schur product,
	we define $U * V$ to be the code in $\mathbb{F}_q^n$ generated by $$\{\boldsymbol{u} * \boldsymbol{v} \mid \boldsymbol{u} \in U~\text{and}~\boldsymbol{v} \in V\}.$$
	

	$\bullet$  For $a=\left ( a_{1},a_{2},\ldots,a_{n}\right )\in \mathbb{F}_{q}^{n}$, the support of $a$ is defined as $$\mathrm{Supp\left ( a \right )}=\left \{ i\mid a_{i}\ne 0, i=1,2,\ldots,n \right \},$$ and the support of a linear code $\mathcal{C}$ of $\mathbb{F}_q^n$ is defined to be $$\mathrm{Supp(\mathcal{C})}=\left \{ \mathrm{Supp\left ( c \right )}\mid c=\left ( c_{1},c_{2},\ldots ,c_{n} \right) \in \mathcal{C}~and~c\ne \mathbf{0} \right \}.$$ Moreover, if $\mathrm{Supp(\mathcal{C})}=[n]$, then $\mathcal{C}$ is a full-support code.

	\subsection{Error-correcting pairs}
	\begin{definition}\label{D_1}{\rm(\cite{A6}, Definition 4.1)}
		For a $q$-ary linear code $\mathcal{C}$ with length $n$. Let $\mathcal{A}$ and $\mathcal{B}$ be linear codes over $\mathbb{F}_{q^m}$ with length $n$, we call $(\mathcal{A}, \mathcal{B})$ an $\ell$-error-correcting pair for $\mathcal{C}$ if the following properties hold simultaneously, 
		$$
		\text { E.1 }~ \mathcal{A} * \mathcal{B} \subseteq \mathcal{C}^{\perp}, \quad \text{ E.2 }~ d(\mathcal{B}^{\perp})>\ell, \quad \text { E.3 }~ k(\mathcal{A})>\ell, \quad \text{ E.4 }~ d(\mathcal{A})+d(\mathcal{C})>n.
		$$
	\end{definition}

	\begin{lemma}\label{P_1}{\rm(\cite{A25}, Theorem 3.1)}
		Let $\ell \in \mathbb{Z}^{+}$ with $2\leq\ell<\frac{n}{2}-1$. For the q-ary MDS linear code $\mathcal{C}=[n, n-2 \ell-1, 2 \ell+2]$, if $\mathcal{C}$ has an $\ell$-error-correcting pair $(\mathcal{A}, \mathcal{B})$, then the parameters of $\mathcal{A}$ have the following three possibilities, 
		$$
		(1)~[n, \ell+2, n-\ell-1];~~~~~~~~
		(2)~[n, \ell+1, n-\ell];~~~~~~~~
		(3)~[n, \ell+1, n-\ell-1].
		$$
	\end{lemma}
	
	\begin{lemma}\label{P_22}{\rm(\cite{A25}, Theorem 3.2)}
		Let $\ell \in \mathbb{Z}^{+}$ with $2\leq\ell<\frac{n}{2}-1$. If $\mathcal{C}$ is an $[n, n-2 \ell-$ $1, 2 \ell+2]$ MDS linear code over $\mathbb{F}_q$ and $(\mathcal{A}, \mathcal{B})$ is an $\ell$-error-correcting pair for $\mathcal{C}$ over $\mathbb{F}_{q^m}$, where $\mathcal{A}=[n, \ell+2, n-\ell-1]_{q^m}$, then 
		
		$(1)$ $\mathcal{C}$ is a GRS code.\\
		Moreover, if $q^m>\max \left\{\left(\begin{array}{l}n \\ i\end{array}\right) \mid 1 \leq i \leq \ell\right\}$, then
		
		$(2)$ $\mathcal{A}$ and $\mathcal{B}$ are both $G R S$ codes with the same evaluation-point sequence as that of $\mathcal{C}$ and $\mathcal{B}=(\mathcal{A} * \mathcal{C})^{\perp}$.
	\end{lemma}
	
	\begin{lemma}\label{P_23}{\rm(\cite{A32}, Theorem 3.1)}
		Let $\ell \in \mathbb{Z}^{+}$ with $2\leq\ell<\frac{n}{2}-1$.
		 If $\mathcal{C}$ is an $[n, n-2 \ell-$ $1, 2 \ell+2]$ MDS linear code over $\mathbb{F}_q$ and $(\mathcal{A}, \mathcal{B})$ is an $\ell$-error-correcting pair for $\mathcal{C}$ over $\mathbb{F}_{q^m}$, where $\mathcal{A}=[n, \ell+1, n-\ell]_{q^m}$.
		 Then for $d(\mathcal{B}^{\perp}) > \ell+1 $, 
		
		$(1)$ $\mathcal{C}$ is a GRS code.\\
		Moreover, if $q^m>\max \left\{\left(\begin{array}{l}n \\ i\end{array}\right) \mid 1 \leq i \leq \ell\right\}$, then
		
		$(2)$ $\mathcal{A}$ and $\mathcal{B}$ are both $G R S$ codes with the same evaluation-point sequence as that of $\mathcal{C}$ and $\mathcal{B}=(\mathcal{A} * \mathcal{C})^{\perp}$.
	\end{lemma}
	
		\subsection{Punctured codes}
	Let $\boldsymbol{x} \in \mathbb{F}_q^n$ and $I=\left\{i_1, \ldots, i_m\right\} \subseteq [n]$ with $1 \leq i_1<\cdots<i_m \leq n$, we define $\boldsymbol{x}_I=\left(x_{i_1}, \ldots, x_{i_m}\right) \in \mathbb{F}_q^m$, namely, $\boldsymbol{x}$ is restricted to the coordinates indexed by $I$. Let $\bar{I}=[n] \backslash I$ be the complement set of $I$ in $[n]$. The punctured code is defined as the following. For more details, see \cite{A11}.
	
	\begin{definition}\label{D_2}
		{\rm({\rm\cite{A11}}, Punctured code)} Let $\mathcal{C}$ be an $[n, k, d]_q$ linear code and $I=\left\{i_1, \ldots, i_m\right\} \subseteq [n]$. The punctured code of $\mathcal{C}$ is defined as
		$$
		\mathcal{C}_I=\left\{c_{\bar{I}} \mid c \in \mathcal{C}\right\}, 
		$$
		which is an $\left[n-m, k\left(\mathcal{C}_I\right), d\left(\mathcal{C}_I\right)\right]_q$ linear code with $d-m \leq d\left(\mathcal{C}_I\right) \leq d$ and $k-m \leq k\left(\mathcal{C}_I\right) \leq k$. Moreover, if $m<d$, then $k\left(\mathcal{C}_I\right)=k$.
	\end{definition}
	
	The following lemma is necessary for the proof of Remark \ref{R_21}.
	
	\begin{lemma}\label{L_1}{\rm(\cite{A6}, Lemma 5.3)}
		Let $\mathcal{C}$ be an $[n, k]_q $ MDS linear code and $I$ be a subset of $[n]$ with $m$ elements such that $n-m \geq k$. Then $\mathcal{C}_I$ is also an MDS linear code with parameters $[n-m, k]_q$.
	\end{lemma}

	\subsection{GRS codes and TGRS codes}
	Now, let us recall the definitions of the GRS and the TGRS code, for more details, see [\citealp{A7},\citealp{A9},\citealp{A11},\citealp{A8}].
	
	\begin{definition}{\rm(\cite{A11})}
		Let $\boldsymbol{\alpha}=\left(\alpha_1, \ldots, \alpha_n\right) \in \mathbb{F}_q^n$ with $\alpha_i \neq \alpha_j(i \neq j)$ and $\boldsymbol{v}=\left(v_1, \ldots, v_n\right) \in$ $\left(\mathbb{F}_q^*\right)^n$. The GRS code with length $n$ and dimension $k$ is defined as
		$$
		G R S_k(\boldsymbol{\alpha}, \boldsymbol{v})=\left\{\left(v_1 f\left(\alpha_1\right), \ldots, v_n f\left(\alpha_n\right)\right) \mid f(x) \in \mathbb{F}_q[x],~\operatorname{deg}(f(x)) \leq k-1\right\}.
		$$
		The $n$-tuple $\boldsymbol{\alpha}$ is called the evaluation-point sequence. It is well-known that $G R S_k(\boldsymbol{\alpha}, \boldsymbol{v})$ is an MDS linear code with parameters $[n, k, n-k+1]_q$ and the corresponding dual code is also a GRS code with the same evaluation-point sequence.
	\end{definition} 
	
	\begin{definition}{\rm(\cite{A7})}
		Let $k, t \in \mathbb{Z}^{+}, h \in \mathbb{N}$ with $0 \leq h<k \leq q$, and $\eta \in \mathbb{F}_q^*$. Define the set of $(k, \eta, t, h)$-twisted polynomials by
		$$
		\mathcal{V}_{k, \eta, t, h}=\left\{f(x)=\sum_{i=0}^{k-1} a_i x^i+\eta a_h x^{k-1+t} \mid a_i \in \mathbb{F}_q,~0 \leq i \leq k-1\right\},
		$$
		which is a $k$-dimensional $q$-ary linear subspace. We call $t$ the twist and $h$ the hook, if $(t,h)=(1,k-1)$, then we abbreviate $\mathcal{V}_{k, \eta, t, h}$ as $(+)-\mathcal{V}_{k, \eta}$.
	\end{definition}
	
	\begin{definition}\label{D_4}{\rm(\cite{A7})}
		Let $\boldsymbol{\alpha}=\left(\alpha_1, \ldots, \alpha_n\right) \in \mathbb{F}_q^n$ with $\alpha_i \neq \alpha_j(i \neq j)$ and $\boldsymbol{v}=\left(v_1, \ldots, v_n\right) \in$ $\left(\mathbb{F}_q^*\right)^n$. The TGRS code with length $n$ and dimension $k$ is defined as
		$$
		\mathcal{C}_k(\boldsymbol{\alpha}, \boldsymbol{v}, \eta, t, h)=\left\{\left(v_1 f\left(\alpha_1\right), \ldots, v_n f\left(\alpha_n\right)\right) \mid f(x) \in \mathcal{V}_{k, \eta, t, h}\right\} .
		$$
		If $(t,h)=(1,k-1)$, then we abbreviate the TGRS code $\mathcal{C}_k(\boldsymbol{\alpha}, \boldsymbol{v}, \eta, t, h)$ as $(+)-\mathcal{C}_k(\boldsymbol{\alpha}, \boldsymbol{v}, \eta)$.
		
	\end{definition}
	
	\begin{lemma}{\rm(\cite{A9}, Lemma 2.6)}\label{L_29}
		For $\eta \in \mathbb{F}_q^*$, $\boldsymbol{\alpha}=\left(\alpha_1, \ldots, \alpha_n\right) \in \mathbb{F}_q^n$ with $\alpha_i \neq \alpha_j(i \neq j)$ and $\boldsymbol{v}=\left(v_1, \ldots, v_n\right) \in$ $\left(\mathbb{F}_q^*\right)^n$. Let
		$$S_{k,+}\left ( \alpha \right )=\left \{ \sum_{i\in I}\alpha _{i}\mid \forall I\subsetneq [n]~and ~\left | I \right |=k \right \},$$
		then we have
		
		$(1)$ $(+)-\mathcal{C}_k(\boldsymbol{\alpha}, \boldsymbol{v}, \eta)$ is MDS if and only if $-\eta ^{-1}\in \mathbb{F}_{q}^{*}\setminus S_{k,+}\left ( \alpha \right )$;
		
		$(2)$ $(+)-\mathcal{C}_k(\boldsymbol{\alpha}, \boldsymbol{v}, \eta)$ is NMDS if and only if $-\eta ^{-1}\in S_{k,+}\left ( \alpha \right )$.
	\end{lemma}

	\subsection{The Schur product of linear codes}
	For convenience, in this section, we suppose that $\mathcal{A}, \mathcal{B} \subseteq \mathbb{F}_{q^m}^n$ are linear codes.
	\begin{lemma}\label{L_22}{\rm(\cite{A3}, Proposition 2.3)}
		If $\mathcal{A}$, $\mathcal{B}$ and $\mathcal{C}$ are linear codes of length $n$ over $\mathbb{F}_q$ such that $\mathcal{A} * \mathcal{B} \subseteq \mathcal{C}^{\perp}$, $d\left(\mathcal{A}^{\perp} \right)>a>0$ and $d\left(\mathcal{B}^{\perp} \right)>b>0$, then $$d\left(\mathcal{C}\right) \geq a+b.$$
	\end{lemma}
	
	\begin{proposition}\label{P_2}{\rm({\rm\cite{A23}}, Product Singleton Bound)}
		Let $\mathcal{A}$ and $\mathcal{B}$ be linear codes, then
		$$
		d(\mathcal{A} * \mathcal{B}) \leq \max \{1, n-(k(\mathcal{A})+k(\mathcal{B}))+2\}.
		$$
	\end{proposition}
	
	Recall that the pair $(\mathcal{A}, \mathcal{B})$ of  $q^m$-linear codes is a Product  MDS (in short, PMDS) pair \cite{A22} if
	$$
	2 \leq d(\mathcal{A} * \mathcal{B})=n-k(\mathcal{A})-k(\mathcal{B})+2.
	$$
	
	\begin{lemma}\label{L_20}{\rm(\cite{A6}, Proposition 7.3)}
		Let $(\mathcal{A}, \mathcal{B})$ be a PMDS pair of $\mathbb{F}_{q}$-linear codes in $\mathbb{F}_{q}^n$. Futher assume that $$k(\mathcal{A})+k(\mathcal{B})<n,$$ then $\mathcal{A}$, $\mathcal{B}$ and $\mathcal{A*B}$ are MDS codes.
	\end{lemma}
	
	\begin{lemma}\label{L_21}{\rm(\cite{A6}, Proposition 7.4)}
		Let $(\mathcal{A}, \mathcal{B})$ be a PMDS pair of $\mathbb{F}_{q}$-linear codes in $\mathbb{F}_{q}^n$. Futher assume that $$k(\mathcal{A})+k(\mathcal{B})<n,$$ and $k(\mathcal{A}), k(\mathcal{B}) \geq 2$. Then, $\mathcal{A}$, $\mathcal{B}$ and $\mathcal{A*B}$ are GRS codes with the common evaluation-point sequence.
	\end{lemma}
	
	
	\begin{proposition}\label{P_3}{\rm(\cite{A22}, Theorem 7)}
		Let $\mathcal{A}$ and $\mathcal{B}$ be full-support codes. If (at least) one of $\mathcal{A}$ and $\mathcal{B}$ is MDS, then
		$$
		k(\mathcal{A} * \mathcal{B}) \geq \min \{n, k(\mathcal{A}) + k(\mathcal{B})-1\} .
		$$
	\end{proposition}
	
	\begin{remark}\label{R_21}
	If $\mathcal{C}$ is an MDS linear code with minimal distance $d \geq 2$, then $\mathcal{C}$ is a full-support code.
	\end{remark}
	
	{\bf Proof.} Suppose that $\mathcal{C}=[n,k,n-k+1]$ is an MDS linear code with $d=n-k+1 \geq2$. Note that $n-1 \geq k$, thus for any $I\subseteq [n] $ with $|I|=1$, $C_I$ is also an MDS linear code with parameters $[n-1,k,n-k]$ by Lemma \ref{L_1}.
	
	If $C$ is not a full-support code, then there exists some $I\subseteq [n]$  with $|I|=1$ and $C_I=[n-1,k,n-k+1]$, which is a contradiction.  $\hfill\Box$\\
	
	
	\begin{lemma}\label{L_10}{\rm(\cite{A22}, Lemma 8)}
		Let $\mathcal{A}$ and $\mathcal{B}$ be MDS codes with
		$$
		k(\mathcal{A} * \mathcal{B})=k(\mathcal{A}) + k(\mathcal{B})-1,
		$$
		then $\mathcal{A} * \mathcal{B}$ is MDS.
	\end{lemma}

		\section{ The $\ell$-error-correcting pair of the  NMDS linear code with minimal distance $2\ell+1$}
		It is well-known that if an NMDS linear code $\mathcal{C}$ with minimal distance $2\ell+1$ has an $\ell$-error-correcting pair $(\mathcal{A}, \mathcal{B})$, then the paramaters of $\mathcal{A}$ have 6 cases given in the following Lemma \ref{L_31}. Firstly, we give an example for that the parameters of $\mathcal{A}$ is the 1st case. Moreover, basing on Product Singleton Bound, we give several necessary conditions for that the parameters of $\mathcal{A}$ is the 4th, 2nd, 5th or 1st case, respectively. 
		
	
		\begin{lemma}\label{L_31}{\rm(\cite{A31}, Theorem 3.5)}
		Let $\ell \in \mathbb{Z}^+$ with $2\leq \ell < \frac{n}{2}-1$. If the q-ary NMDS linear code $\mathcal{C}=[n, n-2 \ell-1, 2 \ell+1]$ has an $\ell$-error-correcting pair $(\mathcal{A}, \mathcal{B})$, then the parameters of $\mathcal{A}$ are the following $6$ possibilities, 
		\begin{align}
	&(A.1) ~[n, \ell+1, n-\ell]; \qquad \quad \notag 
	&&(A.2) ~[n, \ell+2, n-\ell-1]; \quad \notag 
	&&&(A.3)~ [n, \ell+1, n-\ell-1]; \notag \\
	&(A.4) ~[n, \ell+3, n-\ell-2]; \quad \notag 
	&&(A.5) ~[n, \ell+2, n-\ell-2]; \quad \notag 
	&&&(A.6) ~[n, \ell+1, n-\ell-2].  \notag
		\end{align}
	\end{lemma}
	
	$\bullet$ For the case $(A.1)$ of Lemma \ref{L_31}, the example is as follows.
	\begin{example}
		Let $q=37$, $\alpha=(0,1,2,\cdots,8,9)\in \mathbb{F}_{37}^{10}$ and $\mathcal{C}$ be the TGRS code over $\mathbb{F}_{37}$ with the generator matix 
		$$G_{\mathcal{C}}= \begin{pmatrix}
			\mathbf{1} \\
			\alpha  \\
			\alpha^2+6\alpha^3
		\end{pmatrix}_{3\times 10},$$
		then we have $k=3$, $\eta=6$ and $S_{3,+}\left ( \alpha  \right ) =\left \{ 3,4,\cdots ,24 \right \}  $, i.e., $-\eta^{-1}=-6^{-1}=6\in S_{3,+}\left ( \alpha  \right )$. Thus $\mathcal{C}$ is an NMDS linear code with parameters $[10, 3, 7]_{37}$ by Lemma \ref{L_29}.
		
		By taking $\mathcal{A}$ and $\mathcal{B}$ be MDS codes over $\mathbb{F}_{37}$ with the parity check matrix and generator matrix as follows, respectively,
		$$H_{\mathcal{A}}=\begin{pmatrix}
			\mathbf{1} \\
			\alpha  \\
			\alpha^2 \\
			\alpha^3 \\
			\alpha^4 \\
			\alpha^5
		\end{pmatrix}_{6\times 10}  ~~~~~~~~\text{and}~~~~~~~~G_{\mathcal{B}}=\begin{pmatrix}
			\mathbf{1} \\
			\alpha \\
			\alpha^2
		\end{pmatrix}_{3\times 10}. $$
		It's easy to show that $(\mathcal{A}, \mathcal{B})$ is a $3$-error-correcting pair for $\mathcal{C}$ and $\mathcal{A}=[10, 4, 7]_{37}.$ 
	\end{example}

		The following theorem demonstrates that the case $(A.4)$ of Lemma \ref{L_31} can not be satisfied when $2 \leq \ell <\frac{n-3}{2}$.

		\begin{theorem}
			Let $\ell \in \mathbb{Z}^+$ with $2\leq \ell < \frac{n}{2}-1$. If the q-ary NMDS linear code $\mathcal{C}=[n, n-2 \ell-1, 2 \ell+1]$ has an $\ell$-error-correcting pair $(\mathcal{A}, \mathcal{B})$ with $\mathcal{A}=[n, \ell+3, n-\ell-2]_{q^m}$, then $n$ is odd and  
			$\mathcal{C}=[n, 2, n-2]$.
		\end{theorem}
		
		{\bf Proof.}  Let $\ell \in \mathbb{Z}^+$ with $2\leq \ell < \frac{n-3}{2}$. 
		Since $(\mathcal{A}, \mathcal{B})$ is an $\ell$-error-correcting pair for $\mathcal{C}$, thus $d\left(\mathcal{B}^{\perp}\right) \geq \ell+1$ and so $k(\mathcal{B}) \geq \ell \geq 2$. On the other hand, since $\mathcal{A}* \mathcal{C} \subseteq \mathcal{B}^{\perp}$,  $d\left(\mathcal{A}^{\perp}\right) =\ell+4 >\ell+3>0$ and $d\left(\mathcal{C}^{\perp}\right) =n-2\ell-1 > n-2\ell-2>0$, thus $d\left(\mathcal{B} \right) \geq n-\ell+1$ by Lemma \ref{L_22}, i.e., $k(\mathcal{B}) \leq \ell$. Hence 
		$k(\mathcal{B})= \ell$.
		
		 Since $\mathcal{A}*\mathcal{B}\subseteq \mathcal{C}^{\bot}$ and $\mathcal{C}$ is NMDS, thus $d(\mathcal{A}*\mathcal{B})\geq d(\mathcal{C}^{\bot})=n-2\ell-1 \geq 2$. Note that $d(\mathcal{A}*\mathcal{B})\leq max \left\{1, n-\left(k(\mathcal{A})+k(\mathcal{B})\right)+2 \right\}=n-2\ell-1 $, therefore we have
		$$2 \leq d(\mathcal{A}*\mathcal{B})=n-2\ell-1 =  n-\left(k(\mathcal{A})+k(\mathcal{B})\right)+2,$$ hence 
		$(\mathcal{A}, \mathcal{B})$ is a PMDS pair by Proposition \ref{P_2}.
		
		Note that $2\leq \ell < \frac{n-3}{2}$, thus $k(\mathcal{A})+k(\mathcal{B})=2\ell+3<n$, i.e., $\mathcal{A}*\mathcal{B}$ is an MDS linear code with minimal distance $n-2\ell-1$ by Lemma \ref{L_20}.
		Hence $$k(\mathcal{A}*\mathcal{B})=2\ell+2>2\ell+1=k(\mathcal{C}^{\perp}),$$ which contradicts with $\mathcal{A}*\mathcal{B}\subseteq \mathcal{C}^{\perp}$.
		
		From the above, $\ell=\frac{n-3}{2}$, i.e., the parameters of $\mathcal{C}$ is $[n, 2, n-2]$. Furthermore, note that $d(\mathcal{C})=2 \ell+1$ is odd, thus $n$ is odd.
		$\hfill\Box$\\
		
		The following theorem demonstrates that the case ($A.2)$ of Lemma \ref{L_31} also can not be satisfied when $2 \leq \ell <\frac{n-3}{2}$.
		
		\begin{theorem}
		Let $\ell \in \mathbb{Z}^+$ with $2\leq \ell < \frac{n}{2}-1$. If the q-ary NMDS linear code $\mathcal{C}=[n, n-2 \ell-1, 2 \ell+1]$ has an $\ell$-error-correcting pair $(\mathcal{A}, \mathcal{B})$ with $\mathcal{A}=[n, \ell+2, n-\ell-1]_{q^m}$, then $n$ is odd and $\mathcal{C}=[n, 2, n-2]$.
		\end{theorem}
		
		{\bf Proof.} 	Let $\ell \in \mathbb{Z}^{+}$ with $2 \leq \ell<\frac{n-3}{2}$. Since $(\mathcal{A}, \mathcal{B})$ is an $\ell$-error-correcting pair for $\mathcal{C}$, thus $\mathcal{A}* \mathcal{B} \subseteq \mathcal{C}^{\perp}$. Now by Proposition \ref{P_2}, we have
		$$ 2 < n-2\ell-1=d(\mathcal{C}^{\perp}) \leq d(\mathcal{A}*\mathcal{B}) \leq max \left\{1, n-\left(k(\mathcal{A})+k(\mathcal{B})\right)+2 \right\},$$
		note that $k(\mathcal{A})=\ell+2$, thus 
		\begin{equation}\label{eqn_31}
			n-2\ell-1 \leq d(\mathcal{A}*\mathcal{B}) \leq
			 max \left\{1, n-\ell-k(\mathcal{B}) \right\}.
		\end{equation}
		
		Since $(\mathcal{A}, \mathcal{B})$ is an $\ell$-error-correcting pair for $\mathcal{C}$, thus $d\left(\mathcal{B}^{\perp}\right) \geq \ell+1$ and so $k(\mathcal{B}) \geq \ell \geq 2$. On the other hand, since $\mathcal{A}* \mathcal{C} \subseteq \mathcal{B}^{\perp}$,  $d\left(\mathcal{A}^{\perp}\right) =\ell+3 >\ell+2>0$ and $d\left(\mathcal{C}^{\perp}\right) =n-2\ell-1 > n-2\ell-2>0$, thus $d\left(\mathcal{B} \right) \geq n-\ell$ by Lemma \ref{L_22}, i.e., $k(\mathcal{B}) \leq \ell+1$. Hence
		$$ \ell \leq k(\mathcal{B}) \leq \ell+1.$$
		Namely, $k(\mathcal{B})= \ell+1$ or $\ell$.
		
		{\bf Case I.} If $ k(\mathcal{B})=\ell+1$, then $ d(\mathcal{A}*\mathcal{B})=n-2\ell-1=n-\left(k(\mathcal{A})+k(\mathcal{B})\right)+2>2 $ from (\ref{eqn_31}). By Proposition \ref{P_2}, $(\mathcal{A}, \mathcal{B})$ is a PMDS pair, moreover, $\mathcal{A}*\mathcal{B}$ is an MDS linear code with dimension $2\ell +2$ by Lemma \ref{L_20}. Hence $$k(\mathcal{A}*\mathcal{B})=2\ell+2>2\ell+1=k(\mathcal{C}^{\perp}),$$ which contradicts with $\mathcal{A}*\mathcal{B}\subseteq \mathcal{C}^{\perp}$.
		
		{\bf Case II.} If $ k(\mathcal{B})=\ell $, then $n-2\ell-1 \leq d(\mathcal{A}*\mathcal{B}) \leq n-2\ell=n-\left(k(\mathcal{A})+k(\mathcal{B})\right)+2 $ from (\ref{eqn_31}), i.e., $d(\mathcal{A}*\mathcal{B})=n-2\ell$ or $n-2\ell-1$. Moreover, $\mathcal{B}^{\perp}=[n,n- \ell,\ell+1]$ and so $\mathcal{B}$ is a full-support code by Remark \ref{R_21}.
		
		 For $d(\mathcal{A}*\mathcal{B})=n-2\ell$, we know that $\mathcal{A}*\mathcal{B}$ is a PMDS pair by Proposition \ref{P_2}. Moreover, $\mathcal{A}*\mathcal{B}$ is an MDS linear code by Lemma \ref{L_20}, i.e., $k(\mathcal{A}*\mathcal{B})=2\ell+1=k(\mathcal{C}^{\perp})$, note that $\mathcal{A}*\mathcal{B} \subseteq \mathcal{C}^{\perp}$, thus 
		$$\mathcal{A}*\mathcal{B} = \mathcal{C}^{\perp}=[n, 2\ell+1, n-2\ell-1],$$ 
		which contradicts with $d(\mathcal{A}*\mathcal{B})=n-2\ell$. 	
		
		 For $d(\mathcal{A}*\mathcal{B})=n-2\ell-1$, since $\mathcal{A}$ and $\mathcal{B}$ are both full-support codes, by Proposition \ref{P_3} we have $ k(\mathcal{A}*\mathcal{B}) \geq 2\ell+1$. Note that $\mathcal{A}*\mathcal{B}\subseteq \mathcal{C}^{\perp}$, thus $k(\mathcal{A*B}) \leq 2\ell+1$, hence we have $$k(\mathcal{A*B})=2\ell+1=k(\mathcal{A})+k(\mathcal{B})-1,$$ which means that $\mathcal{A*B}$ is an MDS linear code by Lemma \ref{L_10}, i.e., $d(\mathcal{A*B})=n-2\ell$, which contradicts with $k(\mathcal{A}*\mathcal{B})=n-2\ell-1$.
		
		From the above, $\ell=\frac{n-3}{2}$, i.e., the parameters of $\mathcal{C}$ is $[n, 2, n-2]$. Furthermore, note that $d(\mathcal{C})=2 \ell+1$ is odd, thus $n$ is odd.	$\hfill\Box$\\
		

		In the same proof as that of Theorem $3.2$,
		we can get the following corollary, which means that for the case $(A.5)$ of Lemma \ref{L_31}, if $\mathcal{A}$ is a full-support code, then the parameters of both $\mathcal{A*B}$ and $\mathcal{B}^{\perp}$ are determined when $2 \leq \ell <\frac{n-3}{2}$.
		
		\begin{corollary}
		Let $\ell \in \mathbb{Z}^{+}$ with $2 \leq \ell<\frac{n-3}{2}$.
		If the q-ary NMDS linear code $\mathcal{C}=[n, n-2 \ell-1, 2 \ell+1]$ has an $\ell$-error-correcting pair $(\mathcal{A}, \mathcal{B})$ with
		   $\mathcal{A}=[n, \ell+2, n-\ell-2]_{q^m}$ a full-support code, then the parameters of both $\mathcal{B}^{\perp}$ and $\mathcal{A}*\mathcal{B}$ are two possibilities as follows,
		   
		   $(1)$ $\mathcal{B}^{\perp}=[n,n-\ell,\ell+1] \quad and \quad \mathcal{A*B}=\mathcal{C}^{\perp}$;
		   
		   $(2)$ 		   
		   $\mathcal{B}^{\perp}=[n, n-\ell, \ell+1] \quad and \quad \mathcal{A}*\mathcal{B}=[n,2\ell+1,n-2\ell-1]$.
		\end{corollary}

		Next, for the case $(A.1)$ of Lemma \ref{L_31}, the following theorem demonstrates that there are 3 cases for the parameters of both $\mathcal{B}^{\perp}$ and $\mathcal{A*B}$ when $2 \leq \ell<\frac{n-3}{2}$.

		\begin{theorem}
			Let $\ell \in \mathbb{Z}^{+}$ with $2 \leq \ell<\frac{n-3}{2}$.
			If the q-ary NMDS linear code $\mathcal{C}=[n, n-2 \ell-1, 2 \ell+1]$ has an $\ell$-error-correcting pair $(\mathcal{A}, \mathcal{B})$ with
			 $\mathcal{A}=[n, \ell+1, n-\ell]_{q^m}$, then the parameters of both $\mathcal{B}^{\perp}$ and $\mathcal{A}*\mathcal{B}$ are three possibilities as follows,
			
			$(1)$ $\mathcal{B}^{\perp}=[n, n-\ell-1, \ell+1]$;
			
			$(2)$ $\mathcal{B}^{\perp}=[n, n-\ell, \ell+1] \quad and \quad \mathcal{A}*\mathcal{B}=[n,2\ell,n-2\ell+1]$;
			
			$(3)$ $\mathcal{B}^{\perp}=[n, n-\ell, \ell+1] \quad and \quad \mathcal{A}*\mathcal{B}=\mathcal{C}^{\perp}$.
		\end{theorem}
		
		{\bf Proof.} Since $(\mathcal{A}, \mathcal{B})$ is an $\ell$-error-correcting pair for $\mathcal{C}$, thus $\mathcal{A}* \mathcal{B} \subseteq \mathcal{C}^{\perp}$. Now by Proposition \ref{P_2}, we have
		$$ 2 < n-2\ell-1=d(\mathcal{C}^{\perp}) \leq d(\mathcal{A}*\mathcal{B}) \leq max \left\{1, n-\left(k(\mathcal{A})+k(\mathcal{B})\right)+2 \right\},$$
		note that $k(\mathcal{A})=\ell+1$, thus 
		\begin{equation}\label{eqn_32}
			n-2\ell-1 \leq d(\mathcal{A}*\mathcal{B}) \leq
			max \left\{1, n-\ell+1-k(\mathcal{B} )\right\}.
		\end{equation}
		
		Since $(\mathcal{A}, \mathcal{B})$ is an $\ell$-error-correcting pair for $\mathcal{C}$, thus $d\left(\mathcal{B}^{\perp}\right) \geq \ell+1$ and so  $k(\mathcal{B}) \geq \ell \geq 2$. On the other hand, since $\mathcal{A}* \mathcal{C} \subseteq \mathcal{B}^{\perp}$,  $d\left(\mathcal{A}^{\perp}\right) =\ell+2 >\ell+1>0$ and $d\left(\mathcal{C}^{\perp}\right) =n-2\ell-1 > n-2\ell-2>0$, thus $d\left(\mathcal{B} \right) \geq n-\ell-1$ by Lemma \ref{L_22}, i.e., $k(\mathcal{B}) \leq \ell+2$. Hence
		$$ \ell \leq k(\mathcal{B}) \leq \ell+2.$$
		Namely, $k(\mathcal{B})= \ell+2$ or $\ell+1$ or $\ell$.
		
		{\bf Case I.} If $ k(\mathcal{B})=\ell+2$, then $ d(\mathcal{A}*\mathcal{B})=n-2\ell-1=n-\left(k(\mathcal{A})+k(\mathcal{B})\right)+2 >2$ from (\ref{eqn_32}). By Proposition \ref{P_2}, $(\mathcal{A}, \mathcal{B})$ is a PMDS pair, moreover, $\mathcal{A}*\mathcal{B}$ is an MDS linear code with dimension $2\ell +2$ by Lemma \ref{L_20}. Hence $$k(\mathcal{A}*\mathcal{B})=2\ell+2>2\ell+1=k(\mathcal{C}^{\perp}),$$ which contradicts with $\mathcal{A}*\mathcal{B}\subseteq \mathcal{C}^{\perp}$.
		
		{\bf Case II.} If $ k(\mathcal{B})=\ell+1 $, then $n-2\ell-1\leq d(\mathcal{A}*\mathcal{B}) \leq n-2\ell=n-\left(k(\mathcal{A})+k(\mathcal{B})\right)+2 $ from (\ref{eqn_32}), i.e., $d(\mathcal{A}*\mathcal{B})=n-2\ell$ or $n-2\ell-1$. Moreover, it is easy to know that $\mathcal{B}^{\perp}=[n, n-\ell-1, \ell+2]$ or $\mathcal{B}^{\perp}=[n, n-\ell-1, \ell+1]$. 
		
		For $d(\mathcal{A}*\mathcal{B})=n-2\ell$, by Proposition \ref{P_2} and Lemma \ref{L_20}, $\mathcal{A}*\mathcal{B}$ is an MDS linear code, i.e., $\mathcal{A*B}=[n, 2\ell+1, n-2\ell]$, hence $k(\mathcal{A*B})=2\ell+1=k(\mathcal{C}^{\perp})$, we have
		$$\mathcal{A*B}=\mathcal{C}^{\perp}=[n, 2\ell+1, n-2\ell-1],$$
		which contradicts with $d(\mathcal{A}*\mathcal{B})=n-2\ell$. 
		
		For $d(\mathcal{A}*\mathcal{B})=n-2\ell-1$, if $\mathcal{B}^{\perp}=[n, n-\ell-1, \ell+2]$, then $\mathcal{B}$ is a full-support code by Remark \ref{R_21}, now
		 by Proposition \ref{P_3} we have $ k(\mathcal{A}*\mathcal{B}) \geq 2\ell+1$. Note that $\mathcal{A}*\mathcal{B}\subseteq \mathcal{C}^{\perp}$, thus $k(\mathcal{A*B}) \leq 2\ell+1$, hence we have $$k(\mathcal{A*B})=2\ell+1=k(\mathcal{A})+k(\mathcal{B})-1,$$ 
		which means that 
		$\mathcal{A}*\mathcal{B}$ is an MDS linear code from Lemma \ref{L_10}, i.e., $d(\mathcal{A*B})=n-2\ell$, which contradicts with $d(\mathcal{A}*\mathcal{B})=n-2\ell-1$. Hence $\mathcal{B}^{\perp} \ne [n, n-\ell-1, \ell+2]$, i.e.,  $$\mathcal{B}^{\perp}=[n, n-\ell-1, \ell+1].$$ 
		Thus we complete the proof of $(1)$.

		{\bf Case III.} If $ k(\mathcal{B})=\ell $, then $n-2\ell-1 \leq d(\mathcal{A}*\mathcal{B}) \leq n-2\ell+1=n-\left(k(\mathcal{A})+k(\mathcal{B})\right)+2 $ from (\ref{eqn_32}), i.e., $d(\mathcal{A}*\mathcal{B})=n-2\ell+1$ or $n-2\ell$ or $n-2\ell-1$. Moreover, $\mathcal{B}^{\perp}=[n,n- \ell,\ell+1]$ and so $\mathcal{B}$ is a full-suppport code by Remark \ref{R_21}.
		
		For $d(\mathcal{A}*\mathcal{B})=n-2\ell+1$, by Proposition \ref{P_2} and  Lemma \ref{L_20}, $\mathcal{A}*\mathcal{B}$ is an MDS linear code with 
		$$\mathcal{A*B}=[n,2\ell,n-2\ell+1] \quad \text{and} \quad \mathcal{B}^{\perp}=[n,n- \ell,\ell+1].$$
		Thus we complete the proof of $(2)$.
		
		For $d(\mathcal{A}*\mathcal{B})=n-2\ell$, by Proposition \ref{P_3} and $\mathcal{A}*\mathcal{B}\subseteq \mathcal{C}^{\perp}$, we have
		$$ 2\ell \leq k(\mathcal{A}*\mathcal{B}) \leq 2\ell+1,$$
		i.e., $k(\mathcal{A}*\mathcal{B})=2\ell$ or $2\ell+1$. If $k(\mathcal{A}*\mathcal{B})=2\ell=k(\mathcal{A})+k(\mathcal{B})-1$, by Lemma $\ref{L_10}$, we know that $\mathcal{A*B}$ is an MDS linear code with minimal distance $n-2\ell+1$, which contradicts with $d(\mathcal{A*B})=n-2\ell$.
		If $k(\mathcal{A*B})=2\ell+1=k(\mathcal{C}^{\perp})$, then $\mathcal{A}*\mathcal{B}=\mathcal{C}^{\perp}=[n, 2\ell+1,n-2\ell-1]$, which contradicts with $d(\mathcal{A*B})=n-2\ell$.
		
		For $d(\mathcal{A}*\mathcal{B})=n-2\ell-1$, by  Proposition \ref{P_3} and $\mathcal{A}*\mathcal{B}\subseteq \mathcal{C}^{\perp}$, we have
		$$ 2\ell \leq k(\mathcal{A}*\mathcal{B}) \leq 2\ell+1 .$$
		i.e., $k(\mathcal{A}*\mathcal{B})=2\ell$ or $2\ell+1$. By Lemma $\ref{L_10}$, we know that $k(\mathcal{A}*\mathcal{B})=2\ell+1$, and then we can get
		$$\mathcal{A}*\mathcal{B}=\mathcal{C}^{\perp}\quad \text{and} \quad \mathcal{B}^{\perp}=[n, n-\ell, \ell+1].$$
		Thus we complete the proof of $(3)$.
		$\hfill\Box$

	\section{ The $\ell$-error-correcting pair of the NMDS linear code with minimal distance $2\ell+2$}
		
		It is well-known that if an NMDS linear code $\mathcal{C}$ with minimal distance $2\ell+2$ has an $\ell$-error-correcting pair $(\mathcal{A}, \mathcal{B})$, 
		then the paramaters of $\mathcal{A}$ have 10 cases given in the following Lemma \ref{L_41}. Firstly, we give an example for that the parameters of $\mathcal{A}$ is the 1st or 2nd case, respectively. Moreover, basing on Product Singleton Bound, we give several necessary conditions for that the parameters of $\mathcal{A}$ is the 7th, 4th, 8th or 2nd case, respectively. 
		

			\begin{lemma}\label{L_41}{\rm(\cite{A31}, Theorem 3.7)}
		Let $\ell \in \mathbb{Z}^+$ with $2\leq \ell < \frac{n-3}{2}$. If the q-ary NMDS linear code $\mathcal{C}=[n, n-2 \ell-2, 2 \ell+2]$ has an $\ell$-error-correcting pair $(\mathcal{A}, \mathcal{B})$, then the parameters of $\mathcal{A}$ have the following $10$ possibilities, 
		\begin{align}
			&(D.1)~ [n, \ell+1, n-\ell]; ~~\notag 
			&&(D.2)~ [n, \ell+2, n-\ell-1];~~  \notag 
			&&&(D.3) ~[n, \ell+1, n-\ell-1]; \notag \\
			&(D.4)~ [n, \ell+3, n-\ell-2]; ~~ \notag 
			&&(D.5) ~[n, \ell+2, n-\ell-2];  ~~\notag 
			&&&(D.6)~ [n, \ell+1, n-\ell-2];  \\ \notag
			&(D.7)~ [n, \ell+4, n-\ell-3]; ~~\notag 
			&&(D.8) ~[n, \ell+3, n-\ell-3]; ~~\notag 
			&&&(D.9) ~[n, \ell+2, n-\ell-3];  \\ \notag
			&(D.10)~ [n, \ell+1, n-\ell-3].  \notag
		\end{align}
	\end{lemma}
	
	$\bullet$ For the cases $(D.1)$-$(D.2)$ of Lemma \ref{L_41}, the corresponding examples are as follows.
	 \begin{example}
	 	Let $q=37$, $\alpha=(0,1,2,\cdots,8,9)\in \mathbb{F}_{37}^{10}$ and $\mathcal{C}$ be the TGRS code over $\mathbb{F}_{37}$ with the generator matix 
	 	$$G_{\mathcal{C}}= \begin{pmatrix}
	 		\mathbf{1} \\
	 		\alpha  \\
	 		\alpha^2 \\
	 		\alpha^3+6\alpha^4
	 	\end{pmatrix}_{4\times 10} ,$$
	 	then we have $k=4$, $\eta=6$ and $S_{4,+}\left ( \alpha  \right ) =\left \{ 6,\cdots,30 \right \}  $, i.e., $-\eta^{-1}=-6^{-1}=6\in S_{4,+}\left ( \alpha  \right )$. Thus $\mathcal{C}$ is an NMDS linear code with parameters $[10, 4, 6]_{37}$ by Lemma \ref{L_29}.
	 	
	 	By taking $\mathcal{A}$ and $\mathcal{B}$ be MDS codes over $\mathbb{F}_{37}$ with the parity check matrix and the generator matrix, respectively,
	 	 $$H_{\mathcal{A}}=\begin{pmatrix}
	 		\mathbf{1} \\
	 		\alpha  \\
	 		\alpha^2 \\
	 		\alpha^3 \\
	 		\alpha^4 \\
	 		\alpha^5 \\
	 		\alpha^6
	 	\end{pmatrix}_{7\times 10} ~~~~~~~~\text{and}~~~~~~~~G_{\mathcal{B}}=\begin{pmatrix}
	 	\mathbf{1} \\
	 	\alpha \\
	 	\alpha^2 
	 	\end{pmatrix}_{3\times 10}. $$
	 	It's easy to show that $(\mathcal{A}, \mathcal{B})$ is a $2$-error-correcting pair for $\mathcal{C}$ and $\mathcal{A}=[10, 3, 8]_{37}$, which is an example for the case $(D.1)$ of Lemma \ref{L_41}. 
	 	
	 	By taking $\mathcal{A}$ and $\mathcal{B}$ be MDS codes over $\mathbb{F}_{37}$ with the parity check matrix and the generator matrix, respectively,
	 	$$H_{\mathcal{A}}=\begin{pmatrix}
	 		\mathbf{1} \\
	 		\alpha  \\
	 		\alpha^2 \\
	 		\alpha^3 \\
	 		\alpha^4 \\
	 		\alpha^5 
	 	\end{pmatrix}_{6\times 10} ~~~~~~~~\text{and}~~~~~~~~G_{\mathcal{B}}=\begin{pmatrix}
	 		\mathbf{1} \\
	 		\alpha 
	 	\end{pmatrix}_{2\times 10}. $$
	 	It's easy to show that $(\mathcal{A}, \mathcal{B})$ is a $2$-error-correcting pair for $\mathcal{C}$ and $\mathcal{A}=[10, 4, 7]_{37}$, which is an example for the case $(D.2)$ of Lemma \ref{L_41}.  
	 \end{example}

	In the same proof as that of Theorem $3.1$,
	we can get the following theorem, which means that the case $(D.7)$ of Lemma \ref{L_41} can not be satisfied when $2 \leq \ell <\frac{n}{2}-2$.

	\begin{theorem}
		Let $\ell \in \mathbb{Z}^+$ with $2\leq \ell < \frac{n-3}{2}$. If the q-ary NMDS linear code $\mathcal{C}=[n, n-2 \ell-2, 2 \ell+2]$ has an $\ell$-error-correcting pair $(\mathcal{A}, \mathcal{B})$ with $\mathcal{A}=[n, \ell+4, n-\ell-3]_{q^m}$, then $n$ is even and  
		$\mathcal{C}=[n, 2, n-2]$.
	\end{theorem}	

	The following theorem demonstrates that the case $(D.4)$ of Lemma \ref{L_41} also can not be satisfied when $2 \leq \ell <\frac{n}{2}-2$.
	
	\begin{theorem}
		Let $\ell \in \mathbb{Z}^+$ with $2\leq \ell < \frac{n-3}{2}$. If the q-ary NMDS linear code $\mathcal{C}=[n, n-2 \ell-2, 2 \ell+2]$ has an $\ell$-error-correcting pair $(\mathcal{A}, \mathcal{B})$ with $\mathcal{A}=[n, \ell+3, n-\ell-2]_{q^m}$, then $n$ is even and  
		$\mathcal{C}=[n, 2, n-2]$.
	\end{theorem}
	
	{\bf Proof.} Let $\ell \in \mathbb{Z}^+$ with $2\leq \ell < \frac{n}{2}-2$. 
	From $\mathcal{A}* \mathcal{B} \subseteq \mathcal{C}^{\perp}$ and Proposition \ref{P_2}, we have
	$$ 2 < n-2\ell-2=d(\mathcal{C}^{\perp}) \leq d(\mathcal{A}*\mathcal{B}) \leq max \left\{1, n-\left(k(\mathcal{A})+k(\mathcal{B})\right)+2 \right\},$$
	note that $k(\mathcal{A})=\ell+3$, thus 
	\begin{equation}\label{eqn_41}
		n-2\ell-2 \leq d(\mathcal{A}*\mathcal{B}) \leq
		max \left\{1, n-\ell-1-k(\mathcal{B}) \right\}.
	\end{equation}
	
	Since $(\mathcal{A}, \mathcal{B})$ is an $\ell$-error-correcting pair for $\mathcal{C}$, thus $d\left(\mathcal{B}^{\perp}\right) \geq \ell+1$ and so $k(\mathcal{B}) \geq \ell \geq 2$. On the other hand, since $\mathcal{A}* \mathcal{C} \subseteq \mathcal{B}^{\perp}$,  $d\left(\mathcal{A}^{\perp}\right) =\ell+4 >\ell+3>0$ and $d\left(\mathcal{C}^{\perp}\right) =n-2\ell-2 > n-2\ell-3>0$, thus $d\left(\mathcal{B} \right) \geq n-\ell$ by Lemma \ref{L_22}, i.e., $k(\mathcal{B}) \leq \ell+1$. Hence 
	$$ \ell \leq k(\mathcal{B}) \leq \ell+1.$$
	Namely, $k(\mathcal{B})= \ell+1$ or $\ell$.

	{\bf Case I.} If $ k(\mathcal{B})=\ell+1$, then $ d(\mathcal{A}*\mathcal{B})=n-2\ell-2=n-\left(k(\mathcal{A})+k(\mathcal{B})\right)+2 >2 $ from (\ref{eqn_41}). By Proposition \ref{P_2}, $(\mathcal{A}, \mathcal{B})$ is a PMDS pair, moreover, $\mathcal{A}*\mathcal{B}$ is an MDS linear code with dimension $2\ell +3$ by Lemma \ref{L_20}. Hence $$k(\mathcal{A}*\mathcal{B})=2\ell+3>2\ell+2=k(\mathcal{C}^{\perp}),$$ which contradicts with $\mathcal{A}*\mathcal{B}\subseteq \mathcal{C}^{\perp}$.
	
	{\bf Case II.} If $ k(\mathcal{B})=\ell $, then $n-2\ell-2 \leq d(\mathcal{A}*\mathcal{B}) \leq n-2\ell-1=n-\left(k(\mathcal{A})+k(\mathcal{B})\right)+2 $ from (\ref{eqn_41}), i.e., $d(\mathcal{A}*\mathcal{B})=n-2\ell-1$ or $n-2\ell-2$. Moreover, $\mathcal{B}^{\perp}=[n,n- \ell,\ell+1]$ and so $\mathcal{B}$ is a full-support code by Remark \ref{R_21}.
	
	For $d(\mathcal{A}*\mathcal{B})=n-2\ell-1$, by Proposition \ref{P_2} we know that $\mathcal{A}*\mathcal{B}$ is a PMDS pair. Moreover, by Lemma \ref{L_20}, $\mathcal{A}*\mathcal{B}$ is an linear MDS code, i.e., $k(\mathcal{A}*\mathcal{B})=2\ell+2=k(\mathcal{C}^{\perp})$, note that $\mathcal{A}*\mathcal{B} \subseteq \mathcal{C}^{\perp}$, thus 
	$$\mathcal{A}*\mathcal{B} = \mathcal{C}^{\perp}=[n, 2\ell+2, n-2\ell-2],$$ 
	which contradicts with $d(\mathcal{A}*\mathcal{B})=n-2\ell-1$. 
	
	For $d(\mathcal{A}*\mathcal{B})=n-2\ell-2$, since $\mathcal{A}$ and $\mathcal{B}$ are both full-support codes, by Proposition \ref{P_3} we have $ k(\mathcal{A}*\mathcal{B}) \geq 2\ell+2$. Note that $\mathcal{A}*\mathcal{B}\subseteq \mathcal{C}^{\perp}$, thus $k(\mathcal{A*B}) \leq 2\ell+2$, hence we have $$k(\mathcal{A*B})=2\ell+2=k(\mathcal{A})+k(\mathcal{B})-1,$$ which means that $\mathcal{A*B}$ is an MDS linear code by Lemma \ref{L_10}, i.e., $d(\mathcal{A*B})=n-2\ell-1$, which contradicts with $k(\mathcal{A}*\mathcal{B})=n-2\ell-2$.
	
	From the above, $\ell=\frac{n}{2}-2$, i.e., the parameters of $\mathcal{C}$ is $[n, 2, n-2]$. Futhermore, note that $d(\mathcal{C})=2\ell+2$ is even, thus $n$ is even.  $\hfill\Box$\\
	
		
		In the same proof as that of Theorem $4.2$,
		we can get the following corollary, which means that for the case $(D.8)$ of Lemma \ref{L_41}, if $\mathcal{A}$ is a full-support code, then the parameters of both $\mathcal{A*B}$ and $\mathcal{B}^{\perp}$ are determined when $2 \leq \ell <\frac{n}{2}-2$.
		
	
	\begin{corollary}
			Let $\ell \in \mathbb{Z}^{+}$ with $2 \leq \ell<\frac{n}{2}-2$.
		If the q-ary NMDS linear code $\mathcal{C}=[n, n-2 \ell-2, 2 \ell+2]$ has an $\ell$-error-correcting pair $(\mathcal{A}, \mathcal{B})$ with
		$\mathcal{A}=[n, \ell+3, n-\ell-3]_{q^m}$ a full-support code, then the parameters of both $\mathcal{B}^{\perp}$ and $\mathcal{A}*\mathcal{B}$ are two possibilities as follows,
		
		$(1)$ $\mathcal{B}^{\perp}=[n, n-\ell, \ell+1] \quad and \quad \mathcal{A*B}=\mathcal{C}^{\perp}$;
		
		$(2)$ $\mathcal{B}^{\perp}=[n, n-\ell, \ell+1] \quad and \quad \mathcal{A*B}=[n,2\ell+2,n-2\ell-2]$.

	\end{corollary}
	
		Now, for the case $(D.2)$ of Lemma \ref{L_41}, the following theorem demonstrates that there are 3 cases for the parameters of both $\mathcal{B}^{\perp}$ and $\mathcal{A*B}$ when $2 \leq \ell<\frac{n}{2}-2$. 

	\begin{theorem}
		Let $\ell \in \mathbb{Z}^{+}$ with $2 \leq \ell<\frac{n}{2}-2$.
		If the q-ary NMDS linear code $\mathcal{C}=[n, n-2 \ell-2, 2 \ell+2]$ has an $\ell$-error-correcting pair $(\mathcal{A}, \mathcal{B})$ with
		$\mathcal{A}=[n, \ell+2, n-\ell-1]_{q^m}$, then the parameters of both $\mathcal{B}^{\perp}$ and $\mathcal{A}*\mathcal{B}$ are three possibilities as follows,
		
		$(1)$ $\mathcal{B}^{\perp}=[n, n-\ell-1, \ell+1]$;
		
		$(2)$ $\mathcal{B}^{\perp}=[n, n-\ell, \ell+1] \quad and \quad \mathcal{A}*\mathcal{B}=[n,2\ell+1,n-2\ell]$;
		
		$(3)$ $\mathcal{B}^{\perp}=[n, n-\ell, \ell+1] \quad and \quad \mathcal{A}*\mathcal{B}=\mathcal{C}^{\perp}$.
	\end{theorem}
	
	{\bf Proof.} 
	From $\mathcal{A}* \mathcal{B} \subseteq \mathcal{C}^{\perp}$ and Proposition \ref{P_2}, we have
	$$ 2 < n-2\ell-2=d(\mathcal{C}^{\perp}) \leq d(\mathcal{A}*\mathcal{B}) \leq max \left\{1, n-\left(k(\mathcal{A})+k(\mathcal{B})\right)+2 \right\},$$
	note that $k(\mathcal{A})=\ell+2$, thus 
	\begin{equation}\label{eqn_42}
		n-2\ell-2 \leq d(\mathcal{A}*\mathcal{B}) \leq
		max \left\{1, n-\ell-k(\mathcal{B}) \right\}.
	\end{equation}
	
	Since $(\mathcal{A}, \mathcal{B})$ is an $\ell$-error-correcting pair for $\mathcal{C}$, thus $d\left(\mathcal{B}^{\perp}\right) \geq \ell+1$ and so $k(\mathcal{B}) \geq \ell \geq 2$. On the other hand, since $\mathcal{A}* \mathcal{C} \subseteq \mathcal{B}^{\perp}$,  $d\left(\mathcal{A}^{\perp}\right) =\ell+3 >\ell+2>0$ and $d\left(\mathcal{C}^{\perp}\right) =n-2\ell-2 > n-2\ell-3>0$, thus $d\left(\mathcal{B} \right) \geq n-\ell-1$ by Lemma \ref{L_22}, i.e., $k(\mathcal{B}) \leq \ell+2$. Hence 
	$$ \ell \leq k(\mathcal{B}) \leq \ell+2.$$
	Namely, $k(\mathcal{B})= \ell+2$ or $\ell+1$ or $\ell$.

	{\bf Case I.} If $ k(\mathcal{B})=\ell+2$, then $ d(\mathcal{A}*\mathcal{B})=n-2\ell-2=n-\left(k(\mathcal{A})+k(\mathcal{B})\right)+2 >2$ from (\ref{eqn_42}). By Proposition \ref{P_2} and Lemma \ref{L_20}, $\mathcal{A}*\mathcal{B}$ is an MDS linear code with dimension $2\ell +3$. Hence $$k(\mathcal{A}*\mathcal{B})=2\ell+3>2\ell+2=k(\mathcal{C}^{\perp}),$$ which contradicts with $\mathcal{A}*\mathcal{B}\subseteq \mathcal{C}^{\perp}$.
	
	{\bf Case II.} If $ k(\mathcal{B})=\ell+1 $, then $n-2\ell-2 \leq d(\mathcal{A}*\mathcal{B}) \leq n-2\ell-1=n-\left(k(\mathcal{A})+k(\mathcal{B})\right)+2 $ from (\ref{eqn_42}), i.e., $d(\mathcal{A}*\mathcal{B})=n-2\ell-1$ or $n-2\ell-2$. Moreover, it is easy to know that $\mathcal{B}^{\perp}=[n, n-\ell-1, \ell+1]$ or $\mathcal{B}^{\perp}=[n, n-\ell-1, \ell+2]$.
	
	For $d(\mathcal{A}*\mathcal{B})=n-2\ell-1$, by Proposition \ref{P_2} and Lemma \ref{L_20}, $\mathcal{A}*\mathcal{B}$ is an MDS linear code, i.e., $\mathcal{A*B}=[n, 2\ell+2, n-2\ell-1]$, hence $k(\mathcal{A*B})=2\ell+2=k(\mathcal{C}^{\perp})$, we have
	$$\mathcal{A*B}=\mathcal{C}^{\perp}=[n, 2\ell+2, n-2\ell-2],$$
	which contradicts with $d(\mathcal{A}*\mathcal{B})=n-2\ell-1$. 
	
	For $d(\mathcal{A}*\mathcal{B})=n-2\ell-2$, if $\mathcal{B}^{\perp}=[n, n-\ell-1, \ell+2]$, then $\mathcal{B}$ is a full-support code by Remark \ref{R_21},  now from Proposition \ref{P_3} we have $ k(\mathcal{A}*\mathcal{B}) \geq 2\ell+2$. Note that $\mathcal{A}*\mathcal{B}\subseteq \mathcal{C}^{\perp}$, thus $k(\mathcal{A*B}) \leq 2\ell+2$, hence we have $$k(\mathcal{A*B})=2\ell+2=k(\mathcal{A})+k(\mathcal{B})-1,$$ 
	which means that 
	$\mathcal{A}*\mathcal{B}$ is an MDS linear code from Lemma \ref{L_10}, i.e., $d(\mathcal{A*B})=n-2\ell-1$, which contradicts with $d(\mathcal{A}*\mathcal{B})=n-2\ell-2$. Hence $\mathcal{B}^{\perp} \ne [n, n-\ell-1, \ell+2]$, i.e.,  $$\mathcal{B}^{\perp}=[n, n-\ell-1, \ell+1].$$ 
	Thus we complete the proof of $(1)$.
	
	{\bf Case III.} If $ k(\mathcal{B})=\ell $, then $n-2\ell-2 \leq d(\mathcal{A}*\mathcal{B}) \leq n-2\ell=n-\left(k(\mathcal{A})+k(\mathcal{B})\right)+2 $ from (\ref{eqn_42}), i.e., $d(\mathcal{A}*\mathcal{B})=n-2\ell$ or $n-2\ell-1$ or $n-2\ell-2$. Moreover, $\mathcal{B}^{\perp}=[n,n- \ell,\ell+1]$ and so $\mathcal{B}$ is a full-support code by Remark \ref{R_21}. 
	
	 For $d(\mathcal{A}*\mathcal{B})=n-2\ell$, by Proposition \ref{P_2} and Lemma \ref{L_20}, $\mathcal{A}*\mathcal{B}$ is an MDS linear code with $$\mathcal{A*B}=[n,2\ell+1,n-2\ell] \quad \text{and} \quad \mathcal{B}^{\perp}=[n,n- \ell,\ell+1].$$
	Thus we complete the proof of $(2)$.
	
	 For $d(\mathcal{A}*\mathcal{B})=n-2\ell-1$, by Proposition \ref{P_3} and $\mathcal{A}*\mathcal{B}\subseteq \mathcal{C}^{\perp}$, we have
	$$ 2\ell+1 \leq k(\mathcal{A}*\mathcal{B}) \leq 2\ell+2 .$$
	i.e., $k(\mathcal{A}*\mathcal{B})=2\ell+1$ or $2\ell+2$.  
	 If $k(\mathcal{A}*\mathcal{B})=2\ell+1=k(\mathcal{A})+k(\mathcal{B})-1$, by Lemma $\ref{L_10}$, we know that $\mathcal{A*B}$ is an MDS linear code with minimal distance $n-2\ell$, which contradicts with $d(\mathcal{A*B})=n-2\ell-1$.
	If $k(\mathcal{A*B})=2\ell+2=k(\mathcal{C}^{\perp})$, then $\mathcal{A}*\mathcal{B}=\mathcal{C}^{\perp}=[n, 2\ell+2,n-2\ell-2]$, which contradicts with $d(\mathcal{A*B})=n-2\ell-1$.
	
	 For $d(\mathcal{A}*\mathcal{B})=n-2\ell-2$, by  Proposition \ref{P_3} and $\mathcal{A}*\mathcal{B}\subseteq \mathcal{C}^{\perp}$, we have
	$$ 2\ell+1 \leq k(\mathcal{A}*\mathcal{B}) \leq 2\ell+2.$$
	i.e., $k(\mathcal{A}*\mathcal{B})=2\ell+1$ or $2\ell+2$. By Lemma $\ref{L_10}$, we know that $k(\mathcal{A}*\mathcal{B})=2\ell+2$, and so
	$$\mathcal{B}^{\perp}=[n, n-\ell, \ell+1] \quad \text{and} \quad \mathcal{A}*\mathcal{B}=\mathcal{C}^{\perp}.$$
	Thus we complete the proof of $(3)$.
	$\hfill\Box$

	\section{ A brief proof for Lemmas $\ref{P_22}$-$\ref{P_23}$}
	In this section, we provide a brief proof for Lemmas $\ref{P_22}$-$\ref{P_23}$, which means that if an MDS linear code $\mathcal{C}$ with minimal distance $2\ell+2$ has an $\ell$-error-correcting pair $(\mathcal{A},\mathcal{B})$ and the parameters of $\mathcal{A}$ is the 1st or the 2nd case in Lemma \ref{P_1}, then $\mathcal{C}$ is a GRS code.
	
	
	\begin{theorem}
		Let $\ell \in \mathbb{Z}^{+}$ with $2\leq\ell<\frac{n}{2}-1$. If  $\mathcal{C}$ is an $[n, n-2 \ell-$ $1, 2 \ell+2]$ MDS linear code over $\mathbb{F}_{q}$ and $(\mathcal{A}, \mathcal{B})$ is an $\ell$-error-correcting pair for $\mathcal{C}$ over $\mathbb{F}_{q^m}$, where $${\bf I.}~\mathcal{A}=[n, \ell+2, n-\ell-1]_{q^m}~~\text{or}~~ {\bf II.}~\mathcal{A}=[n, \ell+1, n-\ell]_{q^m}~\text{with}~ d(\mathcal{B}^{\perp}) > \ell+1,$$ then we have

		$(1)$ $\mathcal{A}$, $\mathcal{B}$ and $\mathcal{C}$ are both GRS codes with the same evaluation-point sequence;
		
		 $(2)$ $\mathcal{B}=(\mathcal{A} * \mathcal{C})^{\perp}$.
	\end{theorem}

	{\bf Proof.} 
	Firstly, we focus on the parameters of $\mathcal{B}$.
	
	For {\bf I.} Since $(\mathcal{A}, \mathcal{B})$ is an $\ell$-error-correcting pair for $\mathcal{C}$, thus $d\left(\mathcal{B}^{\perp}\right) \geq \ell+1$ and so $k(\mathcal{B}) \geq \ell \geq 2$. On the other hand, since $\mathcal{A}* \mathcal{C} \subseteq \mathcal{B}^{\perp}$,  $d\left(\mathcal{A}^{\perp}\right) =\ell+3 >\ell+2>0$ and $d\left(\mathcal{C}^{\perp}\right) =n-2\ell > n-2\ell-1>0$, thus $d\left(\mathcal{B} \right) \geq n-\ell+1$ by Lemma \ref{L_22}, i.e., $k(\mathcal{B}) \leq \ell$. Hence
	$k(\mathcal{B})=\ell$.
	
	For {\bf II.} From $d(\mathcal{B}^{\perp}) > \ell+1$, we have $k(\mathcal{B}) \geq \ell+1$. Since $(\mathcal{A}, \mathcal{B})$ is an $\ell$-error-correcting pair for $\mathcal{C}$, thus  $\mathcal{A}* \mathcal{C} \subseteq \mathcal{B}^{\perp}$. Note that $d\left(\mathcal{A}^{\perp}\right) =\ell+2 >\ell+1>0$ and $d\left(\mathcal{C}^{\perp}\right) =n-2\ell > n-2\ell-1>0$, thus $d\left(\mathcal{B} \right) \geq n-\ell$ by Lemma \ref{L_22}, i.e., $k(\mathcal{B}) \leq \ell+1$. Hence
	$k(\mathcal{B})=\ell+1$.
	
	From the above, we have $k(\mathcal{A})+k(\mathcal{B})=2\ell+2$ for both cases {\bf I}-{\bf II}. 
	
	Note that $\mathcal{A}*\mathcal{B}\subseteq \mathcal{C}^{\bot}$ and $d(\mathcal{C}^{\bot})=n-2\ell$. Therefore for both cases we have
	$$d(\mathcal{A}*\mathcal{B})\geq d(\mathcal{C}^{\bot})=n-2\ell= max \left\{1, n-\left(k(\mathcal{A})+k(\mathcal{B})\right)+2 \right\},$$
	since $2\leq\ell<\frac{n}{2}-1$, hence 
	$(\mathcal{A}, \mathcal{B})$ is a PMDS pair by Proposition \ref{P_2}. Moreover, we also have $$k(\mathcal{A})+k(\mathcal{B})=2\ell+2<n,$$ thus $\mathcal{A}$, $\mathcal{B}$ and $\mathcal{A}*\mathcal{B}$ are both MDS linear codes by Lemma \ref{L_20}.
	
	Now we prove (1).
	
	(1) In fact, from $k(\mathcal{A})$, $k(\mathcal{B})\geq 2$ for both cases, we know that $\mathcal{A}$, $\mathcal{B}$ and $\mathcal{A*B}$ are GRS codes with the same evaluation-point sequence by Lemma \ref{L_21}, moreover, $$k(\mathcal{A*B})=k(\mathcal{A})+k(\mathcal{B})-1=2\ell+1=k(\mathcal{C}^{\bot}).$$
	Thus $\mathcal{C}^{\bot}=\mathcal{A*B}$ is a GRS code, i.e., $\mathcal{C}$ is also a GRS code with the same evaluation-point sequence as that of $\mathcal{A*B}$, $\mathcal{A}$ and $\mathcal{B}$. 
	
	Next, we prove (2).
	
	(2) Since $\mathcal{A}$ and $\mathcal{C}$ are both GRS codes with the same evaluation-point sequence, thus we have $$k(\mathcal{B}^{\bot})=\left\{\begin{matrix}
		n-\ell=k(\mathcal{A})+k(\mathcal{C})-1=k(\mathcal A*\mathcal C), ~~~~~~~~\text{for {\bf I}},  \\
		n-\ell-1=k(\mathcal{A})+k(\mathcal{C})-1=k(\mathcal A*\mathcal C),~~~~ \text{\text{for {\bf II}}},
	\end{matrix}\right.$$ 
	note that $\mathcal{A*C} \subseteq \mathcal{B}^{\bot}$, thus $(\mathcal B)^{\bot}=\mathcal A*\mathcal{C}$, i.e., $\mathcal{B}=(\mathcal{A} * \mathcal{C})^{\perp}$. 
	
	From the above, we complete the proof.$\hfill\Box$
	
	\begin{remark}
		Compared with the proofs of Lemmas $\ref{P_22}$-$\ref{P_23}$, the proof of Theorem 5.1 does not use the condition $$q^m>\max \left\{\left(\begin{array}{l}n \\ i\end{array}\right) \mid 1 \leq i \leq \ell\right\},$$ so the condition $q^m>\max \left\{\left(\begin{array}{l}n \\ i\end{array}\right) \mid 1 \leq i \leq \ell\right\}$ in Lemmas $\ref{P_22}$-$\ref{P_23}$ can be deleted.
	\end{remark}
	
		Moreover, for an MDS linear code $\mathcal{C}$ with minimal distance $2\ell+1$ or $2\ell+2$ and an $\ell$-error-correcting pair $(\mathcal{A}, \mathcal{B})$, the possible parameters of $\mathcal{B}^{\perp}$ and $\mathcal{A*B}$ are summarized in Table 1.
	
	\begin{table}[htbp]
		\centering
		\begin{threeparttable}
			\scriptsize  
			\renewcommand\arraystretch{1.1}
			\setlength\tabcolsep{0.5pt}
			\setlength{\abovecaptionskip}{3pt}
			\setlength{\belowcaptionskip}{0pt}
			\caption{{\footnotesize The possible parameters of $\mathcal{B}^{\perp}$ and $\mathcal{A*B}$ when the MDS code $\mathcal{C}$ has the $\ell$-ECP $(\mathcal{A}, \mathcal{B})$}}
			
			{\begin{tabular}{|c|l|p{14em}|p{23em}|}  
					\hline
					\centering
					$\ell$ & \quad ~~~~~~~~~~MDS & parameters of \centering{$\mathcal{A}$} &\qquad~~~~ possiblities of $\mathcal{B}^{\perp}$ and $\mathcal{A*B}$\\
					
					\hline
					$1 \leq \ell <\frac{n}{2}-1$ &$\mathcal{C}=[n,n-2\ell,2\ell+1]$   &$\mathcal{A}=[n,\ell+1,n-\ell]$  &$\mathcal{C}$, $\mathcal{A}$ and $\mathcal{B}$ are GRS, $\mathcal{B}=(\mathcal{A*C})^{\perp}$\textsuperscript{[\citealp{A6}]}  \\
					
					\cline{1-4}
					&  &$\mathcal{A}=[n,\ell+2,n-\ell-1]$  &$\mathcal{C}$, $\mathcal{A}$ and $\mathcal{B}$ are GRS, $\mathcal{B}=(\mathcal{A*C})^{\perp}$\textsuperscript{[\citealp{A25}]} \\
					
					\cline{3-4}
					&   &  &$\mathcal{C}$, $\mathcal{A}$ and $\mathcal{B}$ are GRS, $\mathcal{B}=(\mathcal{A*C})^{\perp}$ or\\
					
					&   &$\mathcal{A}=[n,\ell+1,n-\ell]$  & $\mathcal{B}^{\perp}=[n,n-\ell,\ell+1], \mathcal{A*B}=[n,2\ell,n-2\ell+1]$ or\\
					
					$2 \leq \ell <\frac{n}{2}-1$ &$\mathcal{C}=[n,n-2\ell-1,2\ell+2]$   &  &$\mathcal{B}^{\perp}=[n,n-\ell,\ell+1]$, $\mathcal{A*B}=[n,2\ell+1,n-2\ell]$\textsuperscript{[\citealp{A32}]} \\
					
					\cline{3-4}
					&  &  &$\mathcal{B}^{\perp}=[n,n-\ell-1,\ell+1]$, $\mathcal{A*B}=\mathcal{C}^{\perp}$ or\\
					
					&  &$\mathcal{A}=[n,\ell+1,n-\ell-1]$ full\textsuperscript{*}  &$\mathcal{B}^{\perp}=[n,n-\ell,\ell+1]$, $\mathcal{A*B}=\mathcal{C}^{\perp}$ or\\
					
					& &  &$\mathcal{B}^{\perp}=[n,n-\ell,\ell+1]$, $\mathcal{A*B}=[n,2\ell,n-2\ell]$\textsuperscript{[\citealp{A37}]} \\
					\hline
		
			\end{tabular}}
			
			\begin{tablenotes}
				\item[*] “$\mathcal{A}$ full” means that $\mathcal{A}$ is a full-support code.
			\end{tablenotes}
		\end{threeparttable}
	\end{table}
	
	

	\begin{table}[htbp]
		\centering
		\begin{threeparttable}
			\scriptsize  
			\renewcommand\arraystretch{1.1}
			\setlength\tabcolsep{0.5pt}
			\setlength{\abovecaptionskip}{3pt}
			\setlength{\belowcaptionskip}{0pt}
			\caption{{\footnotesize The possible parameters of $\mathcal{B}^{\perp}$ and $\mathcal{A*B}$ when the NMDS code $\mathcal{C}$ has the $\ell$-ECP $(\mathcal{A}, \mathcal{B})$}}
			
			{\begin{tabular}{|c|c|p{14em}|p{23em}|}  
					\hline
					\centering
					$\ell$ & NMDS & parameters of \centering{$\mathcal{A}$} &\qquad~~~~ possiblities of $\mathcal{B}^{\perp}$ and $\mathcal{A*B}$\\
					\hline
					$2 \leq \ell <\frac{n}{2}-1$ &$\mathcal{C}=[n,n-2\ell-1,2\ell+1]$   &$\mathcal{A}=[n,\ell+3,n-\ell-2]$  &$\mathcal{C}=[n,2,n-2]$ with $n$ odd \\
					\cline{3-3}
					& &$\mathcal{A}=[n,\ell+2,n-\ell-1]$   & \\
					\cline{1-4}
					 &  &$\mathcal{A}=[n,\ell+2,n-\ell-2]$ full\textsuperscript{*} &$\mathcal{B}^{\perp}=[n,n-\ell,\ell+1], \mathcal{A*B}=\mathcal{C}^{\perp}$ or\\
					 &   &  &$\mathcal{B}^{\perp}=[n, n-\ell, \ell+1], \mathcal{A}*\mathcal{B}=[n,2\ell+1,n-2\ell-1]$\\

					\cline{3-4}
					&   &  &$\mathcal{B}^{\perp}=[n,n-\ell-1,\ell+1]$ or\\
					
					$2 \leq \ell <\frac{n-3}{2}$ &$\mathcal{C}=[n,n-2\ell-1,2\ell+1]$  &$\mathcal{A}=[n,\ell+1,n-\ell]$  & $\mathcal{B}^{\perp}=[n,n-\ell,\ell+1], \mathcal{A*B}=\mathcal{C}^{\perp}$ or\\
					
					&  &  &$\mathcal{B}^{\perp}=[n,n-\ell,\ell+1]$, $\mathcal{A*B}=[n,2\ell,n-2\ell+1]$ \\
					\cline{2-4}
					&$\mathcal{C}=[n,n-2\ell-2,2\ell+2]$   &$\mathcal{A}=[n,\ell+4,n-\ell-3]$  &$\mathcal{C}=[n,2,n-2]$ with $n$ even \\
					\cline{3-3}
					& &$\mathcal{A}=[n,\ell+3,n-\ell-2]$  & \\
					\hline
					 &  &$\mathcal{A}=[n,\ell+3,n-\ell-3]$ full\textsuperscript{*} &$\mathcal{B}^{\perp}=[n,n-\ell,\ell+1]$, $\mathcal{A*B}=\mathcal{C}^{\perp}$ or\\
					 &  & &$\mathcal{B}^{\perp}=[n, n-\ell, \ell+1], \mathcal{A*B}=[n,2\ell+2,n-2\ell-2]$ \\
					\cline{3-4}
					&  &  &$\mathcal{B}^{\perp}=[n,n-\ell-1,\ell+1]$ or\\
					
					$2 \leq \ell <\frac{n}{2}-2$ &$\mathcal{C}=[n,n-2\ell-2,2\ell+2]$  &$\mathcal{A}=[n,\ell+2,n-\ell-1]$  &$\mathcal{B}^{\perp}=[n,n-\ell,\ell+1]$, $\mathcal{A*B}=\mathcal{C}^{\perp}$ or\\
					
					& &  &$\mathcal{B}^{\perp}=[n,n-\ell,\ell+1]$, $\mathcal{A*B}=[n,2\ell+1,n-2\ell]$\\
					\hline

			\end{tabular}}
			
			\begin{tablenotes}
				\item[*] “$\mathcal{A}$ full” means that $\mathcal{A}$ is a full-support code.
			\end{tablenotes}
		\end{threeparttable}
	\end{table}

	\section{Conclusions}
In this manuscript, we give the following two main results.

$(1)$ The conclusions of Sections 3-4 are summarized in Table 2.

$(2)$ A brief proof for Lemmas $\ref{P_22}$-$\ref{P_23}$ is given.


	

\end{document}